\documentclass{article}
\usepackage{spconf,amsmath,graphicx}
\usepackage{algorithm}
\usepackage{algorithmic}
\usepackage{caption}
\usepackage{graphicx}
\usepackage{multicol}
\usepackage{multirow}
\usepackage{float}
\usepackage{subcaption}
\usepackage{mdwlist}
\usepackage{enumitem}
\usepackage{url}
\title{ED-TTS:  Multi-Scale Emotion Modeling using Cross-Domain Emotion Diarization for Emotional Speech Synthesis}  
\name{Haobin Tang$^{1,2 \dagger}$, Xulong Zhang$^{1 \dagger}$, Ning Cheng$^{1 \ast}$, Jing Xiao$^{1}$, Jianzong Wang$^{1}$ \thanks{$^\dagger$Equal Contribution}
\thanks{$^\ast$Corresponding author: Ning Cheng, chengning211@pingan.com.cn}}
\address{$^{1}$Ping An Technology (Shenzhen) Co., Ltd., China\\$^{2}$University of Science and Technology of China}
\begin{document}
\maketitle

\begin{abstract}
Existing emotional speech synthesis methods often utilize an utterance-level style embedding extracted from reference audio, neglecting the inherent multi-scale property of speech prosody. We introduce ED-TTS, a multi-scale emotional speech synthesis model that leverages Speech \textbf{E}motion \textbf{D}iarization (SED) and Speech Emotion Recognition (SER) to model emotions at different levels. Specifically, our proposed approach integrates the utterance-level emotion embedding extracted by SER with fine-grained frame-level emotion embedding obtained from SED. These embeddings are used to condition the reverse process of the denoising diffusion probabilistic model (DDPM). Additionally, we employ cross-domain SED to accurately predict soft labels, addressing the challenge of a scarcity of fine-grained emotion-annotated datasets for supervising emotional TTS training.
\end{abstract}


\section{Introduction}
Recent researches have shown significant progress in emotional text-to-speech (TTS) thanks to the denoising diffusion probabilistic models (DDPM)~\cite{ho2020denoising,songscore}.
EmoDiff~\cite{guo2022emodiff} uses DDPM with classifier guidance~\cite{DBLP:conf/nips/DhariwalN21} to synthesize controllable and mixed emotion. However, such label guidance will lead to low diversity without manual control and is hard to extend to unseen emotions. EmoMix~\cite{tang2023emomix} uses a pre-trained speech emotion recognition (SER) model that extracts high dimensional emotion embedding from reference audio to condition the reverse process of DDPM.  
\begin{figure*}[t!]   
\centering
\includegraphics[width=0.87\linewidth]{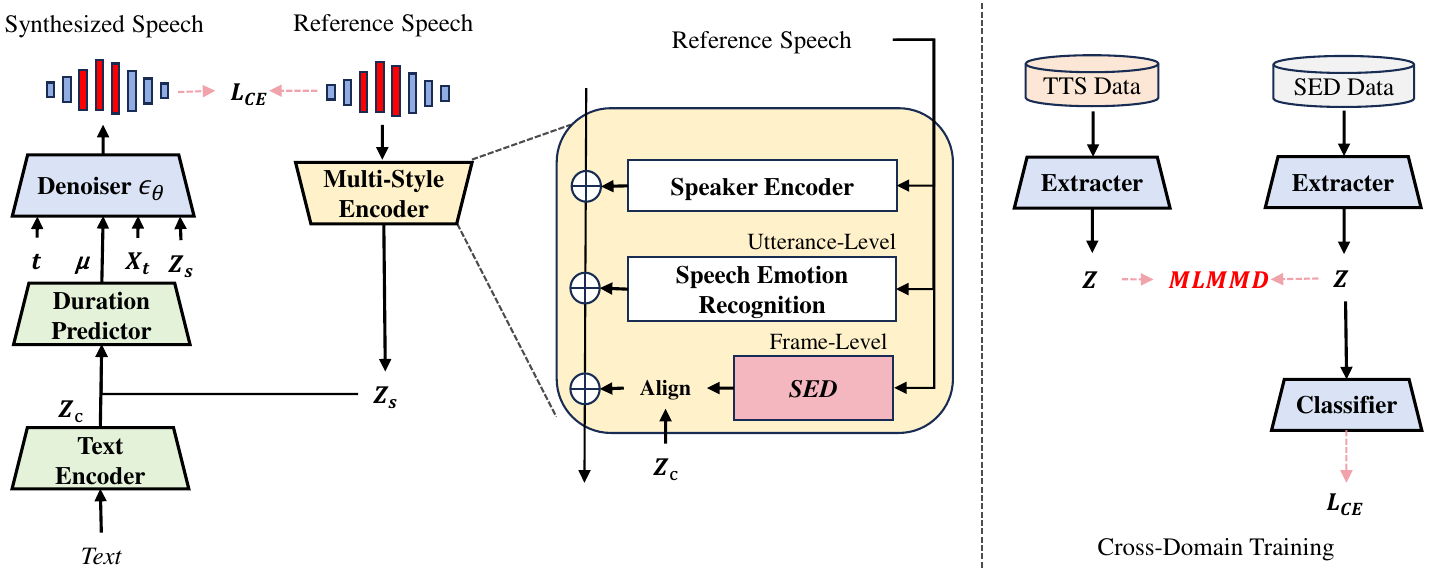}
\caption{The overview of ED-TTS and cross-domain training for SED. The color in waveforms denotes the predicted frame-level emotion labels by SED (e.g. red for non-neutral and blue for neutral). Extracter denotes CNN-based feature encoder of SED.}
\label{Fig1}
% \vspace{em}
\end{figure*}
Such reference-based emotional TTS methods can generate more diverse emotion expression compared to label-based approaches. But the widely used utterance level style embedding fail to capture the multi-scale features of speech style, which span from coarse to fine.
Some fine-grained prosodic expressions, like intonation, are studied. QI-TTS~\cite{tang2023qi} uses a multi-style extractor where the final syllable level style indicates intonation, while the sentence level depicts emotion. 
But sentence level emotion representation can not accurately locating emotion boundaries and fine-grained variations of emotions in the synthesized speech. It’s observed that occasionally, a speaker will stress certain segments of their speech, which makes the emotion more apparent. In that case, the rest of the sentence may sound very neutral. So, we should view emotions expressed in speech as varying speech events that have clear temporal boundaries, instead of the characteristics of the whole speech.

Speech emotion diarization (SED)~\cite{DBLP:journals/corr/abs-2306-12991} is a fine-grained speech emotion recognition task that aims to simultaneously identify the correct emotions and their corresponding boundaries following ``Which emotion appears when?".
To effectively capture the nuances of speech emotion and their boundary, we introduce ED-TTS. This multi-scale approach allows for the modeling of emotions at various levels. ED-TTS is a sequence-to-sequence architecture based on DDPM, with pre-trained SER and SED models that can extract utterance-level and frame-level emotional features. 
Furthermore, we employ SED to address the challenge posed by the scarcity of finely annotated datasets for emotions in emotional TTS. We use the fine-grained soft emotion label predicted by SED on unlabeled TTS dataset to supervise TTS model training. Inspired by Cai et al.~\cite{DBLP:conf/icassp/CaiDWLLM21}, we use cross-domain training to improve the soft label accuracy on TTS datasets by reducing the distribution shift of SED and TTS datasets. 
The main advantages of ED-TTS are:
\begin{enumerate}
    \item ED-TTS is a multi-scale emotional speech synthesis model built on DDPM. It includes two pre-trained components: the utterance-level SER and the frame-level SED. These are designed to identify the category of emotion at the utterance level, and the variation and boundaries of emotion at the frame level, respectively.
    \item ED-TTS further utilizes the SED model to predict frame-level soft emotion labels to supervise TTS model training. Cross-domain training is adopted for improving the performance of SED on TTS dataset.
    \item The results from both subjective and objective evaluation indicate that ED-TTS outshines the baseline models in terms of audio quality and expressiveness.
\end{enumerate}

\section{Proposed Method}
\label{sec:method}
ED-TTS is based on the design of GradTTS~\cite{popov2021grad}, while the multi-scale style encoder use SER as utterance-level extracter and an additional pre-trained SED model for accurately modeling fine-grained emotion feature and their boundaries. We use the extracted multi-scale style embedding to condition the reverse process of DDPM. 
Furthermore, we employ frame-level soft emotion labels predicted by pre-trained cross-domain SED model on TTS dataset to supervise the TTS model training. 

\subsection{Preliminary on Score-based Diffusion Model}
ED-TTS follows GradTTS~\cite{popov2021grad} to apply score-based diffusion model~\cite{songscore} which uses stochastic differential equation (SDE) to TTS. Specifically, it defines a diffusion process which converts any data distribution $X_0$ to terminal distribution $X_T$:
\begin{equation}
\label{eq1}
    d X_{t}=-\frac{1}{2} X_{t} \beta_{t} d t+\sqrt{\beta_{t}} d W_{t}, \quad t \in[0, T]
\end{equation}
where $\beta_{t}$ denotes pre-defined noise schedule and $W_{t}$ is the Wiener process. Above SDE has a corresponding reverse SDE that follows the diffusion process’s reverse trajectory. By solving a discretized version of the reverse time SDE, an ordinary differential equation, GradTTS can generate data $ X_{0}$ from terminal distribution $X_{T}$ as follows:
\begin{equation}
\label{eq2}
    X_{t-\frac{1}{N}}=X_{t}+\frac{\beta_{t}}{N}\left(\frac{1}{2} X_{t}+
    \nabla_{X_{t}}\log p_{t}(X_{t})\right)+\sqrt{\frac{\beta_{t}}{N}} z_{t},
\end{equation}
where $t \in\left\{\frac{1}{N}, \frac{2}{N}, \ldots, 1\right\}$ and $N$ denotes the number of discretized reverse process steps. $z_{t}$ is sampled from standard Gaussian noise. 
But there is an intractable score $\nabla_{X_{t}} \log p_{t}\left(X_{t} \right)$. We can get $X_{t}$ given $X_{0}$ from the distribution derived from Eq.~\eqref{eq1} as $X_{t} \mid X_{0} \sim \mathcal{N}\left(\rho\left(X_{0}, t\right), \lambda(t)\right)$
where $\rho\left(X_{0}, t\right)$ and $\lambda(t)$ have closed form. So that the score $\nabla_{X_{t}} \log p_{t}\left(X_{t} \mid X_{0}\right)=-\lambda(t)^{-1} \epsilon_{t}$, where $\epsilon_{t}$ is the Gaussian noise. To estimate the score a neural network $\epsilon_{\theta}(X_{t}, \mu, t, Z_s)$ is trained using 
\begin{equation}
\label{eq5}
\mathcal{L}_{diff} = E_{x_0, t, Z_s,\epsilon_t}[|| \epsilon_{\theta}(X_{t}, \mu, t, Z_s) +{\lambda(t)}^{-1} \epsilon_t  {||}_2^2]
\end{equation}
where $\mu$ is style and text related Gaussian mean.

\subsection{Multi-Scale Style Encoder}
As shown in the yellow part of Fig.~\ref{Fig1}, we extend the emotion encoder of EmoMix~\cite{tang2023emomix} which contains single SER to multi-scale, to extract the emotion category, emotion variation and emotion boundary information. This module contains the pre-trained SER for utterance-level style features with an additional SED model for frame-level style features. 
We follow the SER~\cite{DBLP:journals/spl/ChenHYZ18} to extract a fixed size embedding from reference speech's mel-spectrogram and its delta, delta-delta coefficients.
The SED~\cite{DBLP:journals/corr/abs-2306-12991} employs pre-trained WavLM~\cite{WavLM}, a modern self-supervised model, followed by a linear classifier. WavLM includes a CNN-based feature encoder followed by transformer blocks and is fine-tuned on the downstream frame-wise SED task. We use the transformer output as our frame-level style embedding. For speaker conditioning we use resemblyzer~\cite{DBLP:conf/icassp/WanWPL18} as our speaker encoder. 

A challenge in fine-grained style conditioning is the alignment of variable-length frame-level prosodic features with input text representations~\cite{DBLP:conf/icassp/ChenR22}. The traditional approach in emotional speech synthesis directly adds style embeddings to text embedding. To align the style representations with the phonetic representations $Z_c$, we adopt the multi-head attention block which aims to reweight content according to the given style learning the alignment between the two modalities. 
The phoneme representations $Z_c$ processed by text encoder is used as query while frame-level style representation is key and value. 
After content-style alignment, the aligned representation is added with utterance-level style embedding and speaker embedding to form multi-scale style embedding $Z_s$. Then $Z_s$ is fed to duration predictor and denoiser to condition the duration modeling and reverse DDPM process.

\subsection{Cross-domain Training of SED}
To minimize the emotion style gap and boundary offsets between reference and synthesized speech. We use SED to predict frame-level soft emotion labels of the unlabelled TTS dataset to supervise ED-TTS training. We train ED-TTS with an additional cross entropy loss which force the synthesized sample have the same frame-level emotion as reference speech.
Since SED is pre-trained on a curated SED dataset which significantly differs from the TTS dataset, we employ domain adaptation techniques to minimize the distribution shift of different datasets.

Kernel-based metric, maximum mean miscrepancy~\cite{DBLP:journals/jmlr/MMD} (MMD), is used to determine the equivalence of two distributions. It has been widely used in domain adaptation tasks and has been validated useful for cross-domain SER within the context of emotional TTS~\cite{DBLP:conf/icassp/CaiDWLLM21}.
Specifically, with the source data $S= \left\{S_1, S_2, \ldots, S_{n_s} \right\}$ and target data  $T= \left\{T_1, T_2, \ldots, T_{n_t} \right\}$ the definition of MMD is
\begin{align}
& \operatorname{MMD}^{2}\left(S, T\right) 
=\left\|\frac{1}{n_{\mathrm{s}}} \sum_{i=1}^{n_{\mathrm{s}}} \phi\left(S_i\right)-\frac{1}{n_{\mathrm{t}}} \sum_{j=1}^{n_{\mathrm{t}}} \phi\left(T_j\right)\right\|_{\mathcal{H}}^{2} \nonumber\\
&= \frac{1}{n_{\mathrm{s}}^{2}} \sum_{i=1}^{n_{\mathrm{s}}} \sum_{j=1}^{n_{\mathrm{s}}} k\left(S_i, S_j\right)+\frac{1}{n_{\mathrm{t}}^{2}} \sum_{i=1}^{n_{\mathrm{t}}} \sum_{j=1}^{n_{\mathrm{t}}} k\left(T_i, T_j\right) \nonumber\\
& \quad-\frac{2}{n_{\mathrm{s}} n_{\mathrm{t}}} \sum_{i=1}^{n_{\mathrm{s}}} \sum_{j=1}^{n_{\mathrm{t}}} k\left(S_i, T_j\right)
\label{eq7}
\end{align}   
where $\phi(\cdot)$ denote a map from data to reproducing kernel hilbert space (RKHS) and $k$ means the Gaussian kernel function.
We divided the source domain and the target domain into different subdomains according to emotion categories to adopt local MMD (LMMD)~\cite{huijuan2023improved} for each subdomains. 
Moreover, we extend LMMD to multi-layer LMMD (MLMMD) which is adopted not only to bottleneck layer but also other CNN layers of the feature encoder part to achieve a more suitable shared feature space. In that case, MLMMD can be expressed as:
\begin{align}
& \operatorname{MLMMD}^{2}\left(S, T\right)\nonumber\\
& =\frac{1}{L\cdot C} \sum_{L=1}^{L}\sum_{C=1}^{C}\left\|\sum_{S_i \in D_{\mathrm{s}}} W_{S_i}^C \phi\left(S_i\right)-\sum_{T_j \in D_{\mathrm{t}}} W_{T_j}^C \phi\left(T_j\right)\right\|_{\mathcal{H}}^{2}
\label{eq8}
\end{align}
where $L$ denotes the count of CNN layers in the SED feature encoder. $C$ is the number of emotion categories. The emotion categories in the source and target domain is mixed or unknown. So we use classification probabilities $W_{S_i}^C$ and $ W_{T_j}^C$ obtained from the pre-trained SER to represent unknown or mixed emotion categories~\cite{tang2023emomix} in the source domain and the target domain respectively. The training of cross-domain SED can be regard as fine-tuning a WavLM model on the down stream SED task and the total loss function for training is:
\begin{equation}
    L=L_{C E}+\lambda L_{M L M M D}
    \label{eq9}
\end{equation}
where $\lambda$ is the weight of MLMMD loss.

\section{Experiments}
\subsection{Dataset}
The SER model is pre-trained on a subset of IEMOCAP~\cite{busso2008iemocap} which contains happy, sad, angry, and neutral emotions. 
The SED model is pre-trained on a curated data~\cite{DBLP:journals/corr/abs-2306-12991} which contain randomly concatenated audio samples from IEMOCAP~\cite{busso2008iemocap}, RAVDESS~\cite{livingstone2018ryerson}, Emov-DB~\cite{adigwe2018emotional}, ESD~\cite{zhou2021seen}, and JL CORPUS~\cite{DBLP:conf/interspeech/JamesTW18}.
We test the cross-domain performance of MLMMD on cross-domain SED tasks on another dataset: Zaion Emotion Dataset (ZED)~\cite{DBLP:journals/corr/abs-2306-12991} which has 180 utterances and 73 speakers derived from emotional YouTube videos. ZED provides discrete emotion labels and emotional segment boundaries for each sample. 
We use a segmented part of BC2013-English audiobook dataset~\cite{BC}, which has about 70 hours and 93k utterances, to train and evaluate ED-TTS. This dataset is read by a single female speaker with expressive style, but without annotations, which fits our task.

\subsection{Experiments Setting}
We train the cross-domain SED model with Adam optimizer under 64 batch size and $10^{-5}$ learning rate setting. 
The score estimation network $\epsilon_\theta$ composed of U-Net and linear attention modules, mirroring those found in GradTTS.
The training of ED-TTS is conducted with 32 batch size and Adam optimizer under a  $10^{-4}$ learning rate for a total of 1 million steps. 
To train the duration predictor, we extract speech-text alignment using Montreal Forced Aligner (MFA)~\cite{mcauliffe2017montreal}. 
For the subsequent experiments, Hifi-GAN~\cite{kong2020hifi} is utilized as the vocoder.

\subsection{Cross-domain SED Results}
To assess the performance of MLMMD in cross-domain SED tasks, we perform experiments on the ZED dataset, measuring the Emotion Diarization Error Rate (EDER) as defined in~\cite{DBLP:journals/corr/abs-2306-12991}. 
EDER metric is specifically designed to accurately assess the temporal alignment between predicted emotion intervals and the actual emotion intervals. We train five models which share the same model structure but use different domain adaptation loss. The weight $\lambda$ in Eq.\eqref{eq9} is set to 0.5.
According to Table 1, the SED-MLMMD model demonstrates a performance enhancement of $3.5\%$ over the SED-base model and $1.7\%$ over the SED-MMD model. This indicates that reducing the distributional gap between two domains enhances the cross-domain SED performance. Extending MMD to Multi-layer Local MMD (MLMMD) contributes to creating a more suitable shared feature space for both the source and target datasets. Moreover, we present the EDER results of Multi-Layer MMD (MMMD) and Local MMD (LMMD) as ablation study for cross-domain SED model.
\begin{table}[htbp]
	\centering
	\caption{ Emotion Diarization Error Rate (EDER) results for cross-domain SED.}
	\scalebox{0.9} {
		\begin{tabular}{lc}
            \hline
            \textbf{Model} & \textbf{EDER $\downarrow$}\\
            \hline
            SED & $31.3  $ \\
            SED-MMD & $29.5 $ \\
            SED-MMMD & $28.2   $\\
            SED-LMMD & $28.6  $ \\
            SED-MLMMD & $\mathbf{27.8}  $ \\
            \hline
            \end{tabular}
	}
	\label{table1}
\end{table}
\subsection{Emotional Speech Evaluation}
\label{section:3.4}
To assess the quality of synthesized speech samples, we compare them with the baseline models:
\begin{enumerate}   
  \item GT and GT (voc.): speech samples from test set and reconstructed speeches usinig vocoder and ground truth mel-spectrogram.
  \item FG-TTS~\cite{DBLP:conf/icassp/ChenR22}: The fine-grained style modeling method based on Transformer TTS.
  \item EmoMix~\cite{tang2023emomix}: A controllable emotional TTS using emotion embedding extracted by a pre-trained SER to condition DDPM.
\end{enumerate}
In subjective evaluation, 25 assessors are tasked with rating 20 speech samples per emotion. They judge the speech quality using the mean opinion score (MOS) and the emotion similarity using the similarity MOS (SMOS), on a scale of 1 to 5.
In objective evaluation, Emotion Reclassification Accuracy (ERA) is used to measure how the synthesized speech fit the frame-level emotion labels of reference speech predicted by SED. Specifically,we reused our pre-trained SED to reclassify the synthesized audio clips and calculated the reclassification accuracy.
Table~\ref{table2} demonstrates that the vocoder’s impact is minimal. ED-TTS outperforms the baseline models in terms of SMOS and ERA by a considerable margin while maintaining MOS scores. These findings highlight ED-TTS’s advantage over the baseline models, attributed to its use of multi-scale emotion modeling and cross-domain training techniques.

\begin{table}[htbp]
	\centering
	\caption{ Evaluation results for emotion synthesis.}
	\scalebox{0.87} {
		\begin{tabular}{lcccc}
            \hline
            \textbf{Model} & \textbf{MOS $\uparrow$}  & \textbf{SMOS $\uparrow$}  & \textbf{ERA $\uparrow$} \\
            \hline
            GT & $4.47 \pm 0.08$ & $4.43 \pm 0.08$ & $-$ \\
            GT (voc.) & $4.40 \pm 0.10$ & $4.38 \pm 0.08$& $0.931$ \\
            \hline
            FG-TTS~\cite{DBLP:conf/icassp/ChenR22} & $3.94 \pm 0.12$ & $3.92 \pm 0.10$& $0.679$ \\       
            EmoMix~\cite{tang2023emomix} & $4.10 \pm 0.10$ & $4.02 \pm 0.08$& $0.623$ \\
            \hline
            ED-TTS & $4.12 \pm 0.08$ & $4.10 \pm 0.12$ & $0.749$ \\
            \hline
            \end{tabular}
	}
	\label{table2}
\end{table}

\subsection{Ablation Study}
To evaluate the impact of techniques used in ED-TTS, including SED utilization, frame-level soft label supervision, and cross-domain training, we conduct ablation studies and present the findings in Table~\ref{table3}. Comparative MOS (CMOS) and comparative ERA (CERA) are utilized to assess the quality and expressiveness of the generated speech. ED-TTS (w/o SED) denotes the ED-TTS model conditioned on single-scale style embedding extracted by SER. The decrease in both quality and reclassification scores shows the significance of modeling emotion style representation at a fine-grained level. Furthermore, the absence of soft label supervision and cross-domain training results in a noticeable decline in CERA, indicating that accurate soft label supervision play a crucial role in guiding ED-TTS to synthesize the correct fine-grained target emotion.
\begin{table}[hbtp]
	\centering
	\caption{CMOS and CERA Results.}
        \begin{tabular}{lcc}
            \hline\hline
            \textbf{Model}& \textbf{CMOS} & \textbf{CERA} \\
            \hline 
            ED-TTS (w/o SED) & -0.07 & -0.12   \\
            ED-TTS (w/o frame label) & -0.04 & -0.08   \\
            ED-TTS (w/o cross domain) & -0.05 & -0.06   \\
            \hline\hline
        \end{tabular}
	\label{table3}
\end{table}

\section{Conclusion}
We propose ED-TTS, a text-to-speech model towards multi-scale style transfer for emotional TTS. We design several techniques to learn the fine-grained emotion variations in speech: 1) ED-TTS employs a multi-scale style encoder to capture and transfer diverse style attributes, encompassing speaker and utterance-level emotional characteristics, as well as nuanced frame-level prosodic representations; 2) ED-TTS uses SED to predict frame-level labels as an auxiliary supervision for TTS model training. Cross-domain training is adopted to SED model for improving the soft emotion label accuracy.

\section{Acknowledgement}
Supported by the Key Research and Development Program of Guangdong Province (grant No. 2021B0101400003) and corresponding author is Ning Cheng. 
% (chengning211@pingan.com.cn). 

\bibliographystyle{IEEEbib}
\bibliography{MSED-TTS}

\end{document}